\documentstyle [12pt]{article}
\newcommand{\be}{\begin{equation}}
\newcommand{\ee}{\end{equation}}
\newcommand{\bea}{\begin{eqnarray}}
\newcommand{\eea}{\end{eqnarray}}
\newcommand{\ba}{\begin{array}}
\newcommand{\ea}{\end{array}}

\begin{document}
\baselineskip = 18 pt
 \thispagestyle{empty}
 \title{
\vspace*{-2.5cm}
\begin{flushright}
\begin{tabular}{c c c c}
\vspace{-0.3cm}
& {\normalsize MPI-Ph/93-37}\\
\vspace{-0.3cm}
& {\normalsize TUM-TH-156/93}\\
\vspace{-0.3cm}
\hspace{0.15cm}
& {\normalsize UAHEP937} \\
\vspace{-0.3cm}
& {\normalsize July 1993}
\end{tabular}
\end{flushright}
\vspace{1.5cm}
 Light Gluinos and Unification of Couplings
\\
 ~\\}
 \author{M. Carena$^{a}$,
 L. Clavelli$^{a,}$\thanks{On leave from the
Dept. of Physics and Astronomy,
University of Alabama, Tuscaloosa.} , D. Matalliotakis$^b$,\\
{}~\\
       H.P. Nilles$^{a,b}$
        $\;$ and  C. E. M. Wagner$^a$\\
 ~\\
$^{(a)}$Max-Planck-Institut f\"{u}r Physik, Werner-Heisenberg-Institut\\
F\"{o}hringer Ring 6, D-80805 M\"{u}nchen , Germany.\\
 ~\\
{}~\\
$^{(b)}$Physik Department, Technische Universit\"at M\"unchen \\
D-85748 Garching, Germany.\\
{}~\\
 }
\date{
\begin{abstract}
We analyze the implications of the light gluino scenario for
the unification of gauge and Yukawa couplings within the
minimal supersymmetric standard model. Within this
scheme all fermionic supersymmetric particles are naturally
light, while the scalar partners of quarks and leptons, together
with the heavy Higgs doublet may be heavy. This implies both a
bound on $\tan\beta < 2.3$, in order to fulfill the experimental
constraints on the chargino  masses, and a strong
correlation between $\sin^2\theta_W(M_Z)$ and $\alpha_3(M_Z)$,
due to the suppression of the supersymmetric threshold corrections
to the low energy values of the gauge couplings.
Assuming the scalar sparticles to be lighter than 10 TeV,
the physical top quark mass is constrained to be
145 GeV $< M_t <$ 210 GeV for $\tan\beta > 1$, while the strong
gauge coupling values,
$0.122 \leq \alpha_3(M_Z) \leq 0.133$, are
in good agreement with the measured LEP ones.
We also show that a relaxation of some of the conventional
assumptions is necessary in order to achieve the radiative
breaking of the electroweak symmetry within the grand unification
scheme.
\end{abstract}}
\maketitle

\newpage
{}~\\
{\bf{Introduction.}}
Unification of couplings
might give hints about the physics at very high
energy scales. The discussion of supersymmetric
Grand Unified Theories (GUTs) has recently attracted much
attention. In fact,
an extrapolation of the measured gauge
couplings within a
minimal supersymmetric extension of the $SU(3)_C\times
{SU(2)_L\times{U(1)_Y}}\ $ standard model leads to a unification
 at a scale $M_X$ of a few $\times10^{16}$GeV, with $M_{SUSY}$, the
supersymmetry
(SUSY) breakdown scale, to be in the 100GeV to TeV energy range,
as expected theoretically. Given the present experimental
uncertainties of the gauge
couplings (especially the strong gauge
coupling  $\alpha_3$), the model is consistent with
unification for a wide range of the parameter space. Assuming
 that at the scale $M_X$ the known forces are contained in a
single grand unified group like SU(5) gives more restrictions.
There one would expect also a unification of certain Yukawa
couplings (like those of the b--quark and the
$\tau$--lepton
in the simplest case) and this, in fact, requires rather
large values for the top quark Yukawa coupling, yielding
predictions
for the mass of the yet undetected top quark which are
remarkably close to the infrared quasi-fixed point value for
this quantity.

In this paper we are going to study  a special class  of such
supersymmetric GUT models in which one parameter, the gaugino
mass $M_{1/2}$ at the unification scale, vanishes (or is
negligible compared to the other parameters which represent
the SUSY breakdown), implying the appearance of a light
gluino in the spectrum.
Although at several instances
it was claimed that the light gluino window is closed, it has to
be accepted that such a particle has not yet been ruled out
experimentally. There has been a thorough discussion of this
issue in the literature and we shall not repeat it here, since
 it is outside the scope of this paper \cite{1}\cite{12}.
One of the recent
motivations to consider the light gluino scenario,
is in relation to the
study of the evolution of the gauge
couplings from low energies up
to the mass of the Z-boson \cite{3}\cite{4}.
A theoretical motivation can be found
in string models where SUSY is broken through
gaugino-condensates, although the prediction of vanishing
gaugino masses within this scheme depends on the special
properties of the potential considered in these works.
In the context of grand  unification such a light gluino
scenario is very appropriate since it gives a restricted range of
 parameter space and therefore enhanced predictive power. In
addition,
the calculation performed in this work can also have more
general application
and stay useful even if such a light gluino would not
exist.

The fact that $M_{1/2}$ is very small combined with the
non-observation of charginos, neutralinos and Higgs bosons
puts severe constraints on the parameter space and primarily
 on $\tan\beta= v_2/v_1$, the ratio of the vacuum
expectation
values of the two Higgs fields. As we shall see, assuming all
supersymmetric particles to have masses lower than 10 TeV,
this then leads
to a very narrow range of values of $\alpha_3$ consistent
with perturbative unification
\be
0.122\le{\alpha_3(M_Z)\le{0.133}}\ ,
\ee
for $1\le{\tan\beta\le{2}}$. A similar
but less restrictive statement
concerning $\alpha_3$ could be made in the more general case,
allowing a heavy gluino, if it were
possible to  restrict the $\tan\beta$
range in a similar way (arguments could come from proton decay
or constraints on the top quark mass but are
still inconclusive). The main difference concerning
the range of allowed values for $\alpha_3(M_Z)$ in the heavy
gluino case however, is due to
to the potentially large
supersymmetric threshold corrections, which, as we shall discuss
below, are much smaller  in the light gluino scenario.
As a second result we find a lower limit on the physical top
quark mass
\be
M_t>145GeV
\label{eq:top}
\ee
when  $\tan\beta\geq$1. It should be pointed out that
(\ref{eq:top}) is mainly a consequence of b--$\tau$
Yukawa coupling unification and also holds in the general case
\cite{7}\cite{10}.
We also find that, in order to achieve the unification of gauge
and Yukawa couplings together with the radiative breaking of
the electroweak symmetry within this scheme, the minimal
supersymmetry breaking conditions at the unification scale should
be modified through, for example, a non--universal scalar mass for
the two Higgs doublets of the theory.

{\bf{Input Data and Constraints.}}
The low energy experimental data and their uncertainties are crucial
 in probing unification. The electroweak parameters at the $Z$--pole
 \cite{8} $M_Z=91.187\pm0.007$ GeV
are presently known to a very good accuracy, the remaining uncertainty
 stemming mainly from the unknown top quark mass. In the modified
minimal subtraction scheme ($\overline{MS}$)  the values that
we use are
\be
\sin^2\theta_W(M_Z) = 0.2324-1.03\cdot10^{-7}GeV^{-2}
\left(M_t^2-(138GeV)^2\right)\pm 0.0003\ ,
\label{eq:sin}
\ee
\be
\frac{1}{\alpha(M_Z)}=127.9\pm0.1\ .
\ee
The quadratic dependence
 of the measured value of $\sin^2\theta_W$ on $M_t$ is explicit in eq.
 (\ref{eq:sin}) and we incorporate this correlation in our analysis.

Unlike the electroweak couplings, the strong gauge coupling
at the $Z$--pole is not so accurately
known. Therefore we do not consider
$\alpha_3(M_Z)$ as an input
in our extrapolation, but rather as
a prediction when we demand gauge
 and bottom to tau Yukawa coupling
 unification. An analysis of the
strong gauge coupling from the existing experimental data
 assuming the presence of light gluinos, has recently  been
carried out
 \cite{5}, indicating that $\alpha_3(M_Z)$ tends to take
significantly higher values.

We take the physical tau mass to be
$M_\tau=1.78$GeV and, neglecting small QED corrections,
we take the running tau mass at
the physical mass $m_{\tau}(M_{\tau})$ to be equal to
the physical mass.
The bottom
 quark mass is less accurately known. We consider for
the physical bottom quark mass the range\cite{12}
\be
4.7GeV\le{M_b\le{5.2}}GeV\ ,
\label{eq:bot}
\ee
and we calculate the running mass at the physical mass through
 the formula
\be
m_b(M_b)=\frac{M_b}{1+\frac{4\alpha_3(M_b)}{3\pi}+
12.4\left(\frac{\alpha_3(M_b)}{\pi}\right)^2}\ ,
\label{eq:cor}
\ee
where two--loop QCD corrections have been taken into account\footnote{
Analogous corrections are considered in the calculation of the
top quark mass. The corresponding formula reads\bea
m_t(M_t)=\frac{M_t}{1+\frac{4\alpha_3(M_t)}
{3\pi}+11\left(\frac{\alpha_3(M_t)}{\pi}\right)^2}\ .
\nonumber
\eea}.

The model we are analysing is significantly more constrained than
 the usual supersymmetric GUTs, and therefore also more predictive
and more easily falsifiable. The most constrained parameter is
 $\tan\beta$. The non--observation of charginos with mass below
half the $Z$ mass imposes an absolute upper bound of 2.3
on $\tan\beta$ \cite{3},\cite{Valle}. In addition,
 experimental lower bounds on the neutralino mass
are very likely to further constrain $\tan\beta$ from
above\footnote{One should be careful when using experimental lower
bounds
in this scenario since the analysis of the data has been carried
 out without taking into account decay modes allowed when gluinos
are light.}.
We take as a reasonable upper bound the value 2, but
we shall also discuss the implications of slightly larger values
of $\tan\beta$.
 Concerning a lower bound on $\tan\beta$, the only
 firm constraint comes through the non--observation of the top quark
 below 108GeV: The requirement of
perturbative consistency of the top quark sector at energy scales
close to  $M_{X}$ implies a lower bound on
 $\tan\beta \geq 0.6$ \cite{16}. However,
 the experimental lower bounds on the light
CP--even Higgs scalar mass make values of $\tan\beta$ lower
than 1 highly improbable, especially, if we constrain
the sparticle masses to be below 1 TeV.
Therefore we will in general consider
$\tan\beta \geq 1$, a requirement naturally appearing in
models with radiative electroweak symmetry breakdown, although
we shall
also discuss how values slightly
below this limit affect our results.

The requirement of b--$\tau$ Yukawa
coupling unification yields predictions for the top quark Yukawa
coupling $h_t$, which are close to its infrared quasi--fixed point
value. So it may
occur that the top quark Yukawa coupling becomes too large at
the high energy scale.
If we want to work consistently in perturbation theory, $h_t$
has to remain perturbative in the whole range of our extrapolation.
To this aim, taking into account the
renormalization group (RG) behaviour of
the top Yukawa, it is enough to constrain $h_t$ at the
unification scale $M_X$. We
require that
\be
\frac{h_t^2(M_X)}{4\pi}<1\ ,
\ee
which is approximately equivalent to the condition that
the two--loop corrections
are less than  $30\%$ of the one--loop contribution.

{\bf{Analysis and Results.}}
In the two--loop
 analysis  we perform, we consider  three distinct
regimes. First we define $M_{scal}$
as the characteristic mass of  squarks,
 sleptons and the heavy Higgs doublet, and we vary it
within the
theoretically and phenomenologically acceptable range
\be
M_Z\le{M_{scal}\le{10TeV}}\ .
\label{eq:sc}
\ee
Thus,
between the unification scale and $M_{scal}$ we use the
 Minimal Supersymmetric Standard Model
 (MSSM) group equations, while below that scale and down to $M_Z$ we
 run the
 Standard Model (SM) renormalization group equations
 with modified $\beta$--function coefficients
 to include the contributions from  gauginos and Higgsinos.
Below $M_Z$ we extrapolate using three-loop QCD including light
gluinos and two-loop QED with chargino
contributions\footnote{We decouple the charginos below
$M_Z$, the exact scale being of no importance for our
results.}. We consider in our analysis the Yukawa couplings
of the third generation quarks and leptons. However, it is worth
mentioning that in the range of $\tan\beta$ we are
examining, the bottom and tau Yukawa coupling contributions
are negligible in comparison to those associated with
the top quark Yukawa coupling.

A few words are neccessary about the decoupling procedure that we
adopt at the various thresholds. At the unification scale we ignore
 possible corrections from splittings in the spectrum of the new
heavy particles introduced, since they are strongly dependent on
the unifying group and require a detailed analysis beyond the
scope of this work. The supersymmetric particles and the
heavy Higgs doublet are decoupled according to the so called
$\theta$-function approximation\footnote{Given the accuracy
of the input data and the two--loop analysis that we pursue,
an attempt to smoothly decouple the heavy modes could have
relevant effects. We choose not to follow this approach
mainly because of the uncertainty concerning smooth decoupling
in the  renormalization schemes we are working with,
but also because we want to make our results readily
comparable with existing analyses.}
: their contribution to the
$\beta$--function coefficients is dropped as soon as we are below
their physical mass and the running couplings are required to
be continuous at the thresholds. Care has been taken to
correctly match the MSSM with the SM Yukawa couplings. The top
quark, the light Higgs and the $SU(2)_L$ gauge bosons are
decoupled by passing from the $SU(3)_C\times{SU(2)_L\times{
U(1)_Y}}$ gauge theory to the effective $SU(3)_C\times{U(1)_{
em}}$ theory. The transition takes place at $M_Z$ and we follow
the conventions of ref. \cite{10}.
When extrapolating between $M_Z$ and the unification scale we
employ the dimensional reduction ($\overline{DR}$) scheme.

We probe unification in the following way: We impose the
conditions
\be
\frac{5}{3}\alpha_1(M_X)=\alpha_2(M_X)=\alpha_3(M_X)=\alpha_X\ ,
\label{eq:un}
\ee
\be
\frac{h_b}{h_\tau}(M_X)=1\ ,
\label{eq:un2}
\ee
where\footnote{Eq. (\ref{eq:un}) is the one--loop exact matching
condition in the $\overline{DR}$ renormalization scheme, if the
heavy particle spectrum is taken degenerate}
$M_X$ is the
unification scale and $\alpha_X$ the gauge coupling
at that scale. We then scan the five--dimensional space defined
by the parameters $M_X$, $\alpha_X$, $h_t(M_X)$, $M_{scal}$
and $\tan\beta$, taking into account the constraints
mentioned previously, and extrapolate
numerically to low energies.
Predicting for each point in the parameter space a set of
values for the low energy quantities $M_t$, $\sin^2\theta_W(M_Z)$,
 $\alpha_{em}(M_Z)$, $\alpha_3(M_Z)$ and $\frac{m_b}{m_\tau}(M_b)$,
 we confirm unification and accept the values of the above
ten parameters as a solution, if eq.(\ref{eq:sin})--(\ref{eq:bot})
 are
satisfied.

The results that we obtain are significantly constrained and highly
correlated. The dominant effect comes from the fact that imposing
b--$\tau$ Yukawa coupling unification drives the top Yukawa coupling to
its infrared quasi-fixed point, thus constraining the top quark
mass to high values (see fig.1). For $M_{scal}$ in the range of
eq. (\ref{eq:sc})
 and $\tan\beta$ between 1 and 2 the physical top quark mass ranges
 in the interval
\be
145GeV<M_t<210GeV\ .
\label{eq:mt}
\ee
Due to  the correlation between the measured value of
$\sin^2\theta_W$ and $M_t$, eq.(\ref{eq:sin}),
and the firm dependence of the predicted value of
$\alpha_3(M_Z)$ on the former, we find solutions only in
the limited range
\be
0.122\le{\alpha_3(M_Z)\le{0.133}}\ ,
\label{eq:as}
\ee
and a tight correlation between
$\alpha_3(M_Z)$ and $M_t$, as can be seen  in fig.2.
There is also a weak
dependence of the predicted values of $\alpha_3(M_Z)$ on the scale
 $M_{scal}$ (see fig.2), which may be understood as follows: Once
the unification condition is imposed, the value of $\alpha_3(M_Z)$
may be given as a function of the electroweak gauge couplings and
the threshold corrections due to the presence of sparticles with
masses above $M_Z$ \cite{8,10}:
\be
\frac{1}{\alpha_3(M_Z)} = \frac{1}{\alpha_3^{SUSY}(M_Z)} +
\frac{19}{28\pi} \ln\left(\frac{T_{SUSY}}{M_Z}\right)\ .
\label{eq:ts}
\ee
There $\alpha_3^{SUSY}(M_Z)$ would be the predicted value for
the strong gauge coupling if the theory were exactly supersymmetric
down to $M_Z$
and $T_{SUSY}$ is an effective scale which characterizes the
supersymmetric threshold corrections to the gauge couplings \cite{10}.
In the light gluino scenario the effective scale $T_{SUSY}$ is
given by
\be
T_{SUSY} = M_Z \left( \frac{M_{scal}}{M_{Z}} \right)^{3/19}
\ee
and hence, for a fixed value of the electroweak gauge couplings,
$\alpha_3(M_Z)$  decreases only slightly as $M_{scal}$ increases.
In fact,
for the range of $M_{scal}$ of eq.(\ref{eq:sc}), $T_{SUSY}$ varies
between the values
\be
M_Z\le{T_{SUSY}\le{200GeV}}\ .
\ee
According to eq.(\ref{eq:ts}) this implies
very small supersymmetric threshold corrections to
the value of $\alpha_3(M_Z)$.

An alternative way of understanding the restricted range of
$\alpha_3$ solutions is to note that the matrix $R$ defined
in Ref.\cite{11}, which relates $\alpha^{-1}_i(M_Z)$ to
$\alpha_X^{-1}$, $\ln\left(M_{scal}/M_Z\right)$ and
$\ln\left(M_X/M_Z\right)$ becomes approximately singular if
only the heavy Higgs, squarks and sleptons are above the
$Z$-boson mass scale, as assumed here. It would be exactly singular
at the one--loop level if the heavy Higgs mass
were also below $M_Z$. In the singular case, $\alpha_3(M_Z)$ would
become independent of $\alpha_X$ and $M_{scal}$, depending then
primarily on $\sin^2\theta_W(M_Z)$. In the language of $T_{SUSY}$
this singular case corresponds to an effective
supersymmetric threshold scale $T_{SUSY} = M_Z$.

We would also like to remark that although there is a weak
dependence of  $\alpha_3(M_Z)$ on $M_{scal}$, this dependence
affects the predicted range for $\alpha_3(M_Z)$ only if
we go to values of $M_{scal}$ above 1TeV. This behaviour is the
result of two counteracting effects: Larger $M_{scal}$ tends to
reduce $\alpha_3(M_Z)$, but at the same time for large $M_{scal}$
 the predicted values of $M_t$ are higher, thus tending to raise
$\alpha_3(M_Z)$. The two effects practically cancel when
$M_{scal}\leq 1$TeV allowing for an almost $M_{scal}$--independent
 range of $\alpha_3(M_Z)$. If we go though to $M_{scal}$ above 1TeV,
we can no longer compensate the decrease in $\alpha_3(M_Z)$ with the
higher values of $M_t$, because for these high values the top
Yukawa coupling develops a Landau singularity making such potential
solutions unacceptable. As a result the upper bound on the range
of $\alpha_3(M_Z)$ decreases when $M_{scal}$ becomes larger than
1TeV. It is worth mentioning that when $M_{scal}\leq 1$TeV $M_t$ is
always found to be smaller than 200GeV.

{}From our analysis
$M_X$  and $\alpha_X$ are  predicted  to lie in the ranges
\be
2.3 \cdot10^{16}GeV<M_X<4.3\cdot10^{16}GeV\ ,
\label{eq:mx}
\ee
\be
23.2\le{\frac{1}{\alpha_X}\le{25.2}}\ .
\label{eq:ax}
\ee

If we now allow $\tan\beta$ to vary in the extended range
\be
0.6\le{\tan\beta\le{2.3}}\ ,
\ee
(keeping in mind that the values below 1 and above 2 are
very likely to be inconsistent with various experimental
lower bounds), the above limits in our results loosen up.
The most dramatic effect exhibits itself in the range
 predicted
for the top quark mass:
\be
104\;GeV\le{M_t\le{212\;GeV}}\ ,
\ee
where the lower limit has significantly decreased responding
 to the low values allowed for $\tan\beta$
 (see fig. 1). The predicted range of values for
the strong gauge coupling is however only slightly enlarged:
\be
0.118\le{\alpha_3(M_Z)\le{0.134}}\ .
\ee
Otherwise the correlations pointed out in the previous
restricted range for $\tan\beta$ also apply in this case.
The lower bound on the
unification scale $M_X$ in eq.(\ref{eq:mx}) goes to the slightly
smaller value $1.9\cdot 10^{16}$ GeV, while the range
of $\alpha_X^{-1}$ is still the one given in eq.(\ref{eq:ax}).

{\bf{Radiative Electroweak Symmetry Breaking.}}
In a recent work \cite{17}, it has been
examined how the requirement of radiative breaking of the
$SU(2)_L\times{U(1)_Y}$ gauge group, when combined with the
current LEP data, constrains the parameter space of the minimal
supergravity--inspired model in the presence of a light gluino.
{}From this analysis it turns out that it is hardly possible to
achieve radiative electroweak breaking in such a model, and the
remaining allowed region of the parameter space might
be experimentally excluded in the near future.
Although we qualitatively agree with this analysis, we find less
stringent limits on the values of the top quark mass consistent with
radiative breakdown. In fact, the range of allowed
values is increased once the uncertainties on $\alpha_3(M_Z)$
are taken into account. In addition, the value of the top
quark mass given in Ref.\cite{17} should be associated with the
running and not the physical mass. Thus, the physical top quark mass
values consistent with radiative electroweak breaking
turn out to be  around 125 GeV,
with $\tan\beta$ between 1.8 and 2.
Even these values of $M_t$ are however inconsistent with unification,
since, as shown in fig.1,
for this range of $\tan\beta$
the predicted top quark mass in a
unified theory is always larger than 180GeV.

One possible way to make radiative electroweak breaking compatible
with unification would be to relax the requirement of exact
bottom to tau Yukawa unification at $M_X$.
In view of the still unknown
 high scale threshold effects and the unclear situation concerning
the generation of fermion masses, this could be considered as a
mild compromise. However it should not escape one's attention that
the difficulty in breaking radiatively the electroweak symmetry
is intrinsic to the vanishing $M_{1/2}$ scenario within the
minimal supersymmetry breaking scheme, independently of unification.
Hence, an alternative way of making the radiative breaking of
$SU(2)_L \times U(1)_Y$ compatible with unification of couplings
would be to relax the universality of the soft supersymmetry
breaking
scalar masses at the high energy scale, through, for
example, non-universal soft supersymmetry breaking
Higgs masses at $M_X$.

In order to understand the above let us review the properties
of the renormalized Higgs mass parameters $m_i^2$
appearing in the potential.
We assume  that at the
grand unification scale, the
squarks and sleptons acquire a common soft supersymmetry
breaking mass  $m_0$ and the Higgs doublets $H_i$ a
breaking mass $m_{H_i}$, while the trilinear term
$A_0$ vanishes.
In the one--loop approximation \cite{0}, the renormalized
values of the Higgs mass parameters $m_1^2$ and $m_2^2$
appearing in the potential are then given by:
\begin{eqnarray}
m_1^2 & \approx & m_{H_1}^2 + \mu^2
\nonumber\\
m_2^2 & \approx & m_{H_2}^2 + \mu^2 - \frac{1}{2}
\left( \frac{ h_t }{ h_t^f } \right)^2 \left(
2 m_0^2 + m_{H_2}^2 \right),
\label{eq:par}
\end{eqnarray}
where $\mu$ is the renormalized SUSY mass parameter appearing
in the superpotential, and $h_t$ and $h_t^f=\sqrt{32\pi\alpha_3/9}$
 are the top quark Yukawa coupling
and its infrared quasi-fixed point value respectively.
 The minimization condition for the potential yields
\be
\tan^2\beta  =  \frac{ m_1^2 + M_Z^2/2}{m_2^2 +M_Z^2/2}\ ,
\label{eq:tanb}
\ee
where we have neglected the radiative correction contributions
which for squarks in the TeV range are of order $M_Z^2$.

{}From our results it follows that unification of gauge and b--$\tau$
Yukawa couplings force the top quark Yukawa coupling to be
at most  10$\%$ away from
 its infrared quasi-fixed point values.
Moreover,
within the light gluino scenario there is an upper bound on
the mass parameter $\mu$ of about $M_Z$. Hence,
in the case of universal soft supersymmetry breaking mass
parameters $m_{H_2} = m_{H_1} = m_0$, and due to the fact that
 the top quark Yukawa
coupling is close to its infrared fixed point value,
 the requirement of radiative breaking
necessarily implies
low values of $m_0 \leq M_Z$, as  can be easily
verified from eqs.(\ref{eq:par}) and (\ref{eq:tanb}).
In addition, it is straightforward to show that both the CP--odd
Higgs mass and the soft SUSY breaking
squark mass terms are also of the order
of or lower than $M_Z$. Thus, the universality of the SUSY
breaking Higgs mass parameter at $M_X$ would imply values for the
lightest CP--even Higgs mass which are below its present
experimental limit and hence the model would be ruled out.
The above conclusion is not preserved if we relax the assumption
of universality of soft SUSY breaking scalar mass terms at the
unification scale.
Indeed, if for example we assume $h_t/h_t^f \simeq 0.9$
(corresponding to a physical top quark mass of about 180GeV
for $\tan\beta\simeq2$),
and $m_{H_2}^2 = 3/2 m_0^2$, while in addition $m_0 \gg M_Z$
in order to avoid problems in the Higgs sector, we obtain
\begin{equation}
\tan^2\beta \simeq \frac{m_{H_1}^2}{0.1 m_{H_2}^2}.
\end{equation}
{}From the above equation, it is easy to verify that, while
varying $m_{H_1}$ from $m_{H_2}/3$  up to
$m_{H_2}/\sqrt{2}$, we can cover values of $\tan\beta$ from
1 up to 2.3. In addition, the SUSY breaking mass parameters
of the supersymmetric partners of the right and left--handed
 top quarks are given by $m_u^2\simeq0.1m_0^2$ and
$m_q^2\simeq0.5m_0^2$ respectively.

The previous results are derived within a one--loop approximation,
where potentially important effects such as squark decoupling
 have been neglected. A more detailed analysis, including two--loop
 corrections to the mass parameters, is necessary if one wants to
make a conclusive statement on the consistency of gauge and Yukawa
coupling unification with radiative electroweak breaking within
this modified soft SUSY breaking scheme.

Observe that, since the Higgs doublets belong to different
representations of $SU(5)$ than the squarks and sleptons, the
relaxation of scalar masses we proposed above is completely
consistent with the $SU(5)$ symmetry. A relaxation of
gaugino mass universality at the grand unification scale
would on the other hand break the $SU(5)$ symmetry of the theory
at $M_X$. One should mention that if this were the case
and, for example, only the partners of the massless gauge
bosons were
below $M_Z$  as was assumed in ref.\cite{3},
significantly lower
strong gauge couplings at $M_Z$ could be obtained.

{\bf{Conclusions.}}
     We have found that SUSY unification with light gluinos
(below the bottom quark mass) constrains the value of the strong
gauge coupling at $M_Z$ to within $\pm 5\%$.  It is
interesting to note that the central value as well as the
allowed range is in good agreement with the current
experimental values from LEP.  However, this result as well
as that of eq.(\ref{eq:mt}) for grand unification in the light gluino
scenario is strongly dependent on the following three
assumptions:
1.   Minimal SUSY particle content.
2.   A common soft supersymmetry breaking
gaugino mass $M_{1/2}$ at the grand unification
scale.
3.   Exact unification of the couplings at a single
scale $M_X$ (see eq.(\ref{eq:un}) and
(\ref{eq:un2})). One should also note that all the
calculations have been done with $\theta$--function
decoupling of the heavy modes. A refinement of this
procedure might have an effect on our results.

Moreover the additional
assumption of radiative breaking of the
$SU(2)_L \times U(1)_Y$
symmetry has been considered. We have shown that a relaxation
of the universality
of the soft supersymmetry scalar masses
associated to the Higgs fields is a possible way to
achieve unification of gauge and b--$\tau$ Yukawa couplings
together with a proper
radiative breaking of the $SU(2)_L \times U(1)_Y$ symmetry
within this scheme.

Finally, we remark that the main difference
between the predictions of the light and
the heavy gluino scenarios is related
to the size of the supersymmetric threshold corrections to
the values of $\alpha_3(M_Z)$. Whereas in the heavy gluino case
these corrections could be as large as  10$\%$, in the light
gluino case, they can not exceed  2$\%$ of the values which
would be obtained if the theory were exactly
supersymmetric down
to $M_Z$. As a consequence, the presence of light gluinos
poses stringent constraints on the allowed range of values
 for $\alpha_3(M_Z)$.

{\bf{Acknowledgments.}}
We would like to thank Stefan Pokorski, F. de Campos
and J. Valle for discussions.
H.P.N. and D.M. are
partially supported by Deutsche
Forschungsgemeinschaft and EC - grant SC1-CT91-0729. L.C. is
partially supported  by the US
Department of Energy under grant no. DE-FG05-84ER40141.
{}~\\
{}~\\

{}~\\
{\bf{FIGURE CAPTIONS}}\\
{}~\\
{}~\\
Fig. 1. Top quark mass as a function of $\tan\beta$, for
$M_{scal} = M_Z$ (solid line)  and $M_{scal} = 2$ TeV
(dashed line). The top quark mass dependence on $\tan\beta$
does not vary for $\sin^2\theta_W$ varying within its
experimental error, eq.(3). \\
{}~\\
Fig. 2. Top quark mass as a function of the strong gauge coupling
for $M_{scal} = M_Z$ (solid lines) and
$M_{scal} = 2$ TeV (dashed lines),
$1 \leq \tan \beta \leq 2$ and
$\sin^2\theta_W$ taking its central value (center), and
its upper  (left) and  lower (right)
experimentally allowed values
at the one-$\sigma$ level, eq.(3). Observe that the $M_S = 2$ TeV
curves are cut at $\alpha_3(M_Z) \simeq 0.13$, due to the
loss of perturbative consistency of the top Yukawa sector of
the theory at $M_X$. \\


\begin{thebibliography}{99}
\bibitem{1}
C.Albajar et al., UA1 Collaboration, Phys. Lett. B198 (1987) 261
\vspace{-0.4cm}
\bibitem{12}
K. Hikasa et al, Particle Data Group,
Phys. Rev. D 45 (1992).
\vspace{-0.4cm}
\bibitem{3} L. Clavelli, Phys. Rev. D46 (1992) 2112,
L. Clavelli, P. Coulter and K. Yuan, Phys. Rev. D 47
(1993) 1973.
\vspace{-0.4cm}
\bibitem{4} M. Jezabek and J. H. K\"uhn, Phys. Lett. B301
(1993) 121.
\vspace{-0.4cm}
\bibitem{7}
H. Arason, D. J. Casta\~no, B. Keszthelyi,
S. Mikaelian, E. J. Piard, P. Ramond and B. D. Wright,
Phys. Rev. Lett. 67 (1991), 2933;
 S. Dimopoulos, L. Hall and S. Raby,
Phys. Rev. Lett. 68 (1992) 1984, Phys. Rev. D45 (1992) 4192;
 S. Kelley, J.L. Lopez and D.V. Nanopoulos,
Phys. Lett. B278 (1992) 140, V. Barger, M.S. Berger
and P. Ohmann, Phys. Rev. D 47 (1993) 1093.
\vspace{-0.4cm}
\bibitem{10}
M. Carena, S. Pokorski and C. Wagner, MPI/Ph 93-10, March 1993, to
appear in Nucl. Phys. B.
\vspace{-0.4cm}
\bibitem{8}
P. Langacker and N. Polonsky, Phys. Rev. D47 (1993)
4028.
\vspace{-0.4cm}
\bibitem{5}
J. Ellis, D. V. Nanopoulos and D. A. Ross,
CERN preprint, CERN TH 6824/93, March 1993.
\vspace{-0.4cm}
\bibitem{Valle}  F. de Campos and J.W.F. Valle, Valencia Univ.
prep. FTUV/93-9, February 1993.
\vspace{-0.4cm}
\bibitem{16}
M. Carena, K. Sasaki and C.E.M. Wagner, Nucl.
Phys. B381 (1992) 66.
\vspace{-0.4cm}
\bibitem{11}
L. Clavelli, Phys. Rev. D45 (1992) 3276
\vspace{-0.4cm}
\bibitem{17}
J. Lopez, D.V. Nanopoulos and X. Wang, CERN
preprint, CERN-TH 6890/93.
\vspace{-0.4cm}
\bibitem{0}
L.E. Ibanez, C.Lopez, C.Munoz, Nucl.Phys. B256 (1985) 218

\end{thebibliography}
\end{document}